\documentclass{aa}
\usepackage{natbib}
\usepackage{graphicx,colortbl}
\usepackage{txfonts}
\usepackage{pdflscape,lscape}
\usepackage{graphicx}
\usepackage{rotating}

\newcommand{\hi}{H\,{\sc i}}

\newcommand{\km}{km\,s$^{-1}$}

\newcommand{\msolar}{M$_{\odot}$}

\newcommand{\degree}{$^{\circ}$}

\definecolor{gold}{rgb}{0.85,.66,0}

\begin{document}

   \title{ {\bf Arp 65 interaction debris: massive \hi\ displacement and star formation }}

   \subtitle{}

   \author{Chandreyee Sengupta, 
          \inst{1}
                    \fnmsep\thanks{}Thomas. C. Scott,\inst{2,3} Sanjaya Paudel,\inst{1} D. J. Saikia,\inst{4,5} K. S. Dwarakanath,\inst{5} B. W. Sohn\inst{1}
          }

   \institute{Korea Astronomy and Space Science Institute, 776, Daedeokdae ro, Yuseong gu, Daejeon, 305-348, Republic of Korea
              \email{csg@kasi.re.kr}
         \and
             \textcolor{black}{Institute of Astrophysics and Space Sciences (IA)}, Rua das Estrelas, 4150-762 Porto, Portugal 
\and
             Centre for Astrophysics Research, University of Hertfordshire, College Lane, Hatfield, AL10 9AB, UK 
 \and
             National Centre for Radio Astrophysics, Tata Institute of Fundamental Research, Pune 411 007, India  
 \and 
              Cotton College State University, Panbazar, Guwahati 781 001, India 
 \and
            Raman Research Institute, Bangalore 560 080, India 
             }

   \date{Received ; accepted }

 
  \abstract
 {\textcolor{black}{ Pre--merger tidal interactions between pairs of galaxies} are known  to induce significant  changes in the morphologies and  kinematics of the  stellar and interstellar medium (ISM) components. Large amounts of gas and \textcolor{black}{stars} are often found to be disturbed or displaced as tidal debris. This debris then evolves, sometimes   forming stars and occasionally forming tidal \textcolor{black}{dwarf} galaxies.  Here we present results from our H{\sc i} study of Arp 65, an interacting pair hosting extended H{\sc i} tidal debris. }
   {In an effort to understand the evolution of tidal debris produced by interacting pairs of galaxies, including in situ star and tidal dwarf galaxy formation, we are mapping   H{\sc i} \textcolor{black}{in}  a sample of interacting galaxy pairs. The Arp 65 pair  is the latest member of this sample to be mapped.}
   {Our  resolved H{\sc i} 21 cm line survey is being carried out using the  Giant Metrewave Radio Telescope (GMRT). We used our H{\sc i} survey data as well as available SDSS optical, \textit{Spitzer } infra-red and \textit{GALEX} UV data to study the evolution of the tidal debris and the correlation of H{\sc i} with the star-forming regions within it.}
   {\textcolor{black}{  In Arp 65 we  see a high impact  pre--merger tidal interaction involving a pair of \textcolor{black}{massive} galaxies (NGC\,90 and NGC\,93) that have  \textcolor{black}{a} stellar mass ratio of  $\sim$ 1:3.}  The interaction,  which probably occurred \textcolor{black}{$\sim$ 1.0 -- 2.5}   $\times$ 10$^8$ yr ago, appears to have displaced a large fraction of the \hi\ in NGC\,90 (including the highest column density \hi)  beyond its optical disk.  We also find extended on-going star formation in the outer disk of NGC\,90. In the major star-forming regions, we find the \hi\ column densities to be $\sim$  4.7 $\times$ 10$^{20}$ cm$^{-2}$ or lower.  But no signature of star formation was found in the highest  column density \hi\ debris SE of NGC\,90. This indicates conditions within \textcolor{black}{the highest density} \hi\ debris remain hostile to star formation and it reaffirms that high H{\sc i} column densities may be a necessary but not sufficient criterion for star formation.}
   {}


\authorrunning{Sengupta et al.}
\titlerunning{H{\sc i} in Arp 65}
\maketitle
\twocolumn

\section{Introduction}

Tidal interactions between pairs of galaxies before their merger are known  to induce significant  changes in the morphologies and  kinematics of the  stellar and  interstellar medium (ISM) components on both  large and  small scales \citep{wielen90,struck05, Chien07}. Observations of pre--merger interacting pairs of galaxies provide the opportunity to investigate the way in which major tidal interactions affect the evolution of galaxies, including the impact on the  rates and location of star formation (SF) as well as  the morphological and kinematic responses of their stellar and  ISM components. The `Tails and Bridges' sample from \cite{smith07}, drawn from the Arp atlas of interacting galaxies, contains a sample of nearby relatively isolated interacting pairs of galaxies. The pairs in  the Smith sample are observed at various stages before merging.  In our resolved  21 cm \hi\  line  studies of Arp pairs with the Giant Metrewave Radio telescope (GMRT) we are focusing on the  impacts of these interactions on the \hi\ morphology and kinematics of late-type galaxies (spirals),  paying  special attention to star formation in the low-density tidal debris and tidal dwarf galaxy (TDG) formation \citep{seng12,seng13,seng14}.

 The Arp\,65 pair  (Fig. 1 -- \textcolor{black}{left}) consists of two large spirals,  NGC\,90 and  NGC\,93. NGC\,90 is a  grand design barred spiral (SABc) with  impressively enhanced spiral arms and long tidal tails to the north-west (NW) and south-east (SE).  NGC\,93 (\textcolor{black}{spiral}) is the more massive of the pair \textcolor{black}{ in terms of stellar mass and is} projected about  3.0$^{\arcmin}$ (64 kpc) to the east of NGC\,90. The galaxies are \textcolor{black}{catalogued} as members of the SRGb063 galaxy group, (Fig. 1 -- \textcolor{black}{right}) n (number of members) = 45, V$_{opt}$ = 5771 \km\ \citep{mahdavi04}. NGC\,90 is projected $\sim$ 3.3$^{\arcmin}$ (70 kpc) to the east of the group's intra-group medium (IGM) centroid based on X--ray observations by  \cite{mahdavi04}. Further basic properties of the pair  are given in Table \ref{table_1}. 

We present here results from our GMRT \hi\ observations of the Arp 65 system. In addition to the \hi\ data from the GMRT, we have used the Sloan Digital Sky Survey (SDSS), \textit{Spitzer} and Galaxy Evolution Explorer (\textit{GALEX}) public archive data and images to understand the relation between the gaseous and  stellar components of the system. The paper is \textcolor{black}{set out}  as follows. Sect. 2 gives details of the \textcolor{black}{GMRT} observations,  with  observational results in Sect. 3.  A discussion follows in Sect. 4 with concluding remarks in Sect. 5. The optical \textcolor{black}{heliocentric} radial velocities of NGC\,90 and NGC\,93 are respectively 5197 \km\ and 5643 \km\ (Hyperleda\footnote{http://atlas.obs-hp.fr/hyperleda/}). Using the average radial velocity (5420 \km) and assuming H$_{0}$ to be 75 km s$^{-1}$ Mpc$^{-1}$,  we estimate the distance to Arp 65 pair to be 73 Mpc. Based on this distance the spatial scale is $\sim$ 21.2 kpc per arcmin.  

\begin{table}
\centering
\begin{minipage}{190mm}
\caption{Properties of the Arp 65 pair}
\label{table_1}
\begin{tabular}{llrr}
\hline
Property\footnote{All data  are from NED, except for V$_{radial(optical)}$ and inclination\\
which are from Hyperleda}&Units&NGC\,90&NGC\,93 \\ 
\hline
V$_{radial(optical)}$&[\km]& 5197$\pm$25  &5643$\pm$20\\
RA&[h:m:s]&00:21:51.4&00:22:03.2\\
DEC&[d:m:s]&+22:24:00.0&+22:24:29.0\\
Distance\footnote{Method  detailed in Sect. 1.}&[Mpc]&  73 & 73 \\
D$_{25}$ major /minor  &[arcmin]&  1.9 x 0.8 & 1.4 x 0.6  \\
D$_{25}$ major /minor& [kpc]& 39.0 x 16.4 & 28.7 x 12.3\\
Inclination& [\degree] & 34.5  & 59.3   \\
Morphology&&SABc&S?  \\
B$_T$& [B band mag]&14.54$\pm$ 0.13   &14.34$\pm$0.13   \\
Stellar mass $M_*$  &[10$^{10}$ \msolar] &  4.7 &  14 \\
\end{tabular}
\end{minipage}
\end{table}



\section{Observations}
\label{obs}
\hi\  21 cm line  observations of Arp 65 were carried out  using  the GMRT on 15 March, 2009. Further details of the observations  are given in Table \ref{table2}. The baseband bandwidth used was 16 MHz for the \hi\   observations giving a velocity resolution of $\sim$27 \km. The  Astronomical Image Processing System ({\tt AIPS}) software package was used for data reduction. Bad data  from malfunctioning antennas and antennas with low gain and/or  radio frequency interference were flagged.   The flux densities are on the scale of \cite{baars77}, with flux density uncertainties of  $\sim$5\%.  After calibration, continuum subtraction in the uv domain was carried out using the {\tt AIPS} tasks  \textsc{uvsub} and \textsc{uvlin}. The task \textsc{imagr} was then used to obtain  the final cleaned  \hi\  image cubes. The integrated \hi\  and  \hi\   velocity field maps were extracted from the cube using the AIPS task \textsc{momnt}. To analyse the  \hi\ structures,  image cubes of different resolutions were produced by applying different `tapers' to the data with varied \textit{u,v} limits.


\begin{table}
\centering
\begin{minipage}{110mm}
\caption{GMRT observation details}
\label{table2}
\begin{tabular}{ll}
\hline


Frequency & 1420.407 MHz \\
Observation date &15th March, 2009 \\
Primary calibrator&3C286\\ 
Phase \textcolor{black}{calibrator} (flux density) &\textcolor{black}{ 0029+349 (1.89 Jy)}  \\
Integration time  & 8.0 hr  \\
Primary beam&24\arcmin ~at 1420.4057 MHz \\
Low-resolution beam & 48.3$^{\prime\prime}$ $\times$ 38.3$^{\prime\prime}$  (PA = -36.3$^{\circ}$) \\
High-resolution beam & 23.4$^{\prime\prime}$ $\times$ 15.2$^{\prime\prime}$  (PA = -29.7$^{\circ}$) \\

rms for low-resolution map  & 1.2 mJy beam$^{-1}$  \\
rms for high-resolution map & 1.0 mJy beam$^{-1}$  \\

RA (pointing centre)& 00$^{\rm h}$ 21$^{\rm m}$ 51.4$^{\rm s}$  \\
DEC (pointing centre)&22$^\circ$ 23$^{\rm m}$ 58.6$^{\rm s}$\\

\hline
\end{tabular}
\end{minipage}
\end{table}

\section{Observational results}
\label{results}
\subsection {\hi\ content and distribution}
\label{results_content}
 The top panel  of  Fig. 2  shows the low-resolution  integrated \hi\ contours (in white) \textcolor{black}{ overlaid} on \textcolor{black}{an} SDSS \textit{r}-band image for Arp\,65.   The blue contours in the figure show the far-ultraviolet (FUV)  \textcolor{black}{emission} from \textit{GALEX}. The bulk of the H{\sc i}  detected in this system  is associated  with the SABc galaxy  NGC\,90. The other member of the pair, NGC\,93  appears to be a more evolved spiral with a redder SDSS \textit{g--i} colour of 1.58  compared to NGC\,90 (SDSS \textit{g--i} =1.4). \hi\ is only  marginally detected in NGC\,93.   The most striking  feature in the low-resolution map is that the  H{\sc i} intensity maximum and the bulk of the \hi\  \textcolor{black}{is} projected  $\sim$ 1 arcmin (21 kpc) to the south-east of the NGC\,90 optical centre and beyond its  optical disk. 


The integrated \hi\ flux density obtained  by \cite{springob05}  from single-dish observations for NGC\,90 was 5.4 Jy \km\, as compared to  4.9 Jy \km\ from our GMRT observations, meaning that the GMRT recovered $\sim$ 90\% of single-dish  \hi\ flux density. Conversion of the single dish  flux density to an \hi\ mass gives   $M_{HI}$ =  6.8 $\times$ 10$^{9}$ M$_\odot$.  We use this  \hi\ mass derived from the single-dish observations  for all our  calculations requiring  the \textcolor{black}{NGC\,90}  \hi\ mass. The single-dish \hi\ flux density for NGC\,93 was 1.46 Jy \km\  \citep{springob05}.  However the  marginal H{\sc i} detection for NGC\,93  in our GMRT map (Fig. 2, top panel),  the Arecibo beam size ($\sim$ 3.5 arcmin) and the coincidence in velocity of flux maxima in the Arecibo spectrum of NGC\,93 (in red) with the high-velocity horn of  the GMRT spectrum (in black) for NGC\,90 \textcolor{black}{(Fig.2, middle right panel)}  leads us to conclude that most  of H{\sc i} flux in the NGC\,93 Arecibo \textcolor{black}{single-dish} spectrum is in fact \hi\ spatially associated  with NGC\,90.


The middle right panel  of Fig. 2 \textcolor{black}{also shows \hi\ spectra for NGC\,90 from both the GMRT (in black) and the single-dish (in blue) observations.  The single-dish spectrum has V$_{HI}$ = 5333 \km\ and W$_{20}$ = 477 \km\  \citep{springob05}. \textcolor{black}{ This W$_{20}$ value is $\sim$ 45 \km\ higher than an estimate of W$_{20}$ based on the M* values in section  \ref{discuss-interaction} and equation 19 in \cite{tf11}.  } }  The GMRT resolved images  and channel maps provide insights into the  \textcolor{black}{enhancement of the } line width. The middle left panel of Fig. 2 shows the velocity field for  \hi\ for  NGC\,90 (colour scale) overlaid with  the  FUV (\textit{GALEX})  contours (in black). The red contour line shows  the  H{\sc i} extent (3 $\sigma$) from the GMRT low-resolution  integrated \hi\ map. West of the NGC\,90 optical centre the velocity field  indicates a more or less regular velocity gradient.  But the eastern part of the disk shows an anomalous velocity behaviour. The separation between high- and low-velocity components in the channel maps (Fig. 5) is indicated by the lack of  \hi\ detections between  5347 \km\ and 5402 \km. A  gas blob appears SE of NGC\,90 at velocities $>$ 5000 \km\ and \textcolor{black}{moves} NW across the optical disk with a velocity increasing  up to $\sim$ 5320 \km. At around 5480 \km\ another blob appears  SE of the NGC\,90 optical disk and remains detected there at velocities up to 5650 \km. Based on the channel maps, we divided the high- and low-velocity components of NGC\,90  at 5350 \km.  The bottom left panel of Fig 2 shows the velocity field of the low-velocity component, overplotted with the total intensity contours \textcolor{black}{(in blue)} of the high-velocity component and the FUV (\textit{GALEX}) contours (in black) defining the edge of the \textcolor{black}{optical} galaxy. The \hi\ velocity field  for low-velocity component \hi\ in NGC\,90 presents \textcolor{black}{as} a disk with  fairly regular rotation and no significant asymmetry. This velocity field shows that the \hi\ velocity at the optical centre of NGC\,90 is $\sim$ 5188 \km\ (V$_{optical}$ = 5197$\pm$25) and its  rotation velocity  is  $\sim$ 100 \km, \textcolor{black}{before adjustment for inclination}. No sign of regular rotation is found in the velocity field for the high-velocity component, referred to as the ``SE high-velocity \hi\ mass'' from now on. This separation of high- and low-velocity gas is also clearly seen in the position-velocity (pv) diagram (Fig. 3). The signal-to-noise ratio (S/N) is poor in the pv diagram, but it successfully shows the gradient in the disk of NGC 90 and the lack of a velocity gradient in the SE high-veloctiy gas mass. About 50\% of the total \hi\ mass of 6.8 $\times$ 10$^{9}$ M$_\odot$ can be attributed to the low-velocity component (NGC\,90) implying $\sim$ 3.4 $\times $10$^{9}$ M$_\odot$  for the SE high-velocity \hi\ mass. We conclude that both the abnormally large \hi\ line width and the perturbed \hi\ morphology of NGC\,90  are largely the result of the superposition of the SE high-velocity \hi\ mass on the gas disk of NGC\,90 that rotates with a lower velocity.



\subsection { \hi\  and star formation beyond the NGC\,90 optical disk }
\label{results_sf}
As the top panel of Fig. 2 shows, NGC\,90 hosts several star forming zones  beyond its optical disk along the western tidal tail and east of the disk. The most prominent, labelled as SF1, SF2 and SF3, are visible in both the SDSS \textit{r}-band and the  FUV (\textit{GALEX}) images. H{\sc i} is detected at the positions of  these star-forming zones, but without any evidence of local enhancement in \hi\ flux density. The  H{\sc i} column densities\footnote{$N(HI)=(1.1\times10^{21}/a \times b)\times\int{S_{\nu}dv}$; where N(HI) is the H{\sc i} column density, a and b are the beam major and minor axes, S$_\nu$ is the flux per channel and dv is the per channel velocity width. The integral is over all the channels with line emission (Fig. 1, top panel).}  in  these star-forming zones \textcolor{black}{are} $\sim$  1.2 -- 4.7 $\times$ 10$^{20}$ cm$^{-2}$. \textcolor{black}{The high- and low-resolution \hi\ maps both show the column density maxima projected at $\sim$ 10 kpc east of the  NGC\,90 optical disk edge (within the SE high-velocity \hi\ mass).} The H{\sc i} column density maxima  from the  \textcolor{black}{high}-resolution map (Fig. 2 bottom right panel)  is $\sim$ \textcolor{black}{7 to 8}  $\times$10$^{20}$ cm$^{-2}$.  There is no evidence of ongoing star formation at  or near the position of H{\sc i} column density maxima \textcolor{black}{from} \textcolor{black}{ the optical,  UV or \textit{Spitzer} 24 $\mu$m}  images.

\begin{table}
\centering
\begin{minipage}{110mm}
\caption{GMRT \hi\ observation results}
\label{table3}
\begin{tabular}{ll}
\hline

\hi\ mass (total) in the Arp 65 system  &  6.8 $\times$ 10$^{9}$ \msolar\ $\pm$ 0.7 $\times$ 10$^{9}$ \msolar\ \\ 
\hi\ mass of the NGC\,90 \textcolor{black}{low-velocity disk}&  3.4 $\times$ 10$^{9}$ \msolar\ $\pm$ 0.4 $\times$ 10$^{9}$ \textcolor{black}{ \msolar} \\
\hi\ mass of the \textcolor{black}{SE high-velocity \hi\ mass }&  3.4 $\times$ 10$^{9}$ \msolar\ $\pm$ 0.4 $\times$ 10$^{9}$\textcolor{black}{ \msolar} \\
\hi\ velocity of NGC\,90 & 5188 $\pm$ 27  \km  \\
\hi\ velocity of the SE debris & 5531 $\pm$ 27  \km \\
\textcolor{black}{ \hi\ rotation velocity of NGC 90} & 100 $\pm$ 27 \km \\
\hline

\end{tabular}
\end{minipage}
\end{table}

\section{Discussion}
In this discussion we consider what the observations to date, including our GMRT \hi\ data, reveal about the interaction between the pair and SF in the tidal debris. 

\label{discuss}
\subsection{Evidence of an NGC\,90/NGC\,93 interaction}
\label{discuss-interaction}
The estimated stellar masses (M*)  for NGC\,90  and NGC\,93 are   $\sim$  4.9 $\times$ 10$^{10}$ \msolar\ and   1.4 $\times$ 10$^{11}$ \msolar\  respectively, based on the parameters from \cite{bell03}  and \cite{blanton03} for their SDSS \textit{r}-band magnitudes and \textit{r-i} colours. The M* values for NGC\,90 (including the tails) and   NGC\,93  derived from the formula in \cite{eskew12} using \textit{Spitzer} 3.6$\mu$m  and 4.5$\mu$m flux densities from \cite{smith07} are  4.4 $\times$ 10$^{10}$ \msolar\ and 1.4 $\times$ 10$^{11}$ \msolar\  respectively. Both methods produce M* values that agree well with each other and give a stellar mass ratio of $\sim$ 1:3.\\
Interacting galaxy models \textcolor{black}{ and observations \citep[e.g.][]{mihos01,oh08,casteels13}} indicate enhanced stellar spiral arms \textcolor{black}{as well as transitory  tidal bridges and}   tails can  result from prograde  high-impact, low-velocity flyby encounters.  Such interactions are expected to produce a long optical tidal tail on the far side and a bridge  joining the pair on the near side   \citep[]{mihos01}. Strong arm enhancement is expected in bulge-dominated galaxies and  a bar in disk-dominated galaxies. A stellar bridge is not visible between the NGC\,90/93 pair, but a bar, \textcolor{black}{ prominent stellar arms  and  a  far side} tidal tail are observed in NGC\,90 \textcolor{black}{(Fig. 1, left panel).}  
 
 \cite{oh08}  used models to study the enhancement by a perturber of the optical spiral arms, tidal bridges  and  tidal tails (together with their visibility time-scales) in a disk galaxy. The baryonic mass of NGC\,90 ($\sim$ 5.4 $\times$ 10$^{10}$ \msolar) is close to that of the Oh model disk galaxy  ($\sim$ 5.2 $\times$ 10$^{10}$ \msolar), therefore it is reasonable to assume that the galaxy properties, including the dynamical times-cale, are sufficiently similar  to make valid comparisons between the two, accepting that the orbital parameters remain poorly constrained compared to the Oh models.  The Oh models use a tidal strength parameter,\\ $S$ = ($\frac{M_p}{M_g}$) ($\frac{R_g}{R_{peri}}$)$^3$ ($\frac{\Delta T}{T}$) (their Eq. 3),\\
   and investigate interactions with  $S$ $< 0.3$. For  NGC\,90 we can say the ratio of perturber to  galaxy mass ($\frac{M_p}{M_g}$) is $\sim$ 3. But we have no direct constraint on the more important ratio of the peri--centre to galaxy radius ($\frac{R_g}{R_{peri}}$) variable or $\frac{\Delta T}{T}$, the angular speed of the perturber relative to stars at the galaxy edge. In the \cite{oh08} models the  tidal tails dissipate rapidly after reaching their most prominent phase 1.4 to 2.5 $\times$ 10$^8$ yr after the interaction. But for S $>$ 0.3 the tail  dissipation time-scale may extend to $\sim$ 1 Gyr and in this case the tail may fragment to form TDGs \citep{barnes92,oh08}. However \textcolor{black}{four} pieces of evidence support an interaction within the times-cales derived from the \cite{oh08} models: (i) \hi\  morphological perturbation signatures from a full merger of galaxies with  total baryonic masses of the same order as NGC\,90, only remain identifiable for a maximum of  $\sim$ 4  $\times$10$^{8}$ yr to  7 $\times$ 10$^{8}$ yr   \citep{holwerda11}. \textcolor{black}{ It seems likely that \hi\ perturbations from a full merger would be of the same order as the pre--merger interaction we observe in NGC\,90. } But because the \hi\ morphology perturbations in NGC\,90 are extreme compared to \textcolor{black}{ the Holwerda \hi\ merger} signature thresholds, \textcolor{black}{ it is reasonable to conclude the NGC\,90} perturbation occurred well within the upper limits of the  \citep{holwerda11} times-cale; (ii) the projected distance between  NGC\,90 and NGC\, 93 is $\sim$ 64 kpc. Assuming the separation velocity is the SRGb0063 group velocity dispersion of $\sim$ 336 \km, we  estimate the shortest possible time  since their closest approach to be $\sim$ 1.9  $\times$ 10$^{8}$ years ago; (iii) \textcolor{black}{the time for a single rotation\footnote{ T$_{rot}$ [Gyr]= 6.1478 r/ V$_{rot}$, where r = the optical radius [kpc] and V$_{rot}$= 0.5 $\Delta$V  [\km]/ sin(i).} of NGC\,90 \textcolor{black}{ is $\sim$ 0.7 $\times$ 10$^{9}$ yr, based on the $\sim$ 180 \km\   \hi\ rotational velocity (inclination adjusted) of the low-velocity component and data from Tab. 2.  \textcolor{black}{A} de--projected  SDSS \textit{g} -- band  image\footnote{based on  PA = 113.4\degree\ and inclination = 34.5\degree\  from HyperLeda} of NGC\,90 }  indicates that \textcolor{black}{the truncated bridge remnant} is now} offset $\sim$ 60\degree\ counter-clockwise from a line joining the \textcolor{black}{optical centres} of NGC\,90 and NGC\,93. \textcolor{black}{ If we assume that the NGC\,90/NGC\,93 bridge formed along an axis joining the two galaxies soon after their closest approach,  as interaction models predict,  then the current $\sim$ 60\degree\ offset  (between a line joining the two galaxies and the bridge remnant) reflects the approximate fraction  the time for  single NGC\,90 disk rotation since the bridge was formed. \textcolor{black}{ This then  implies  a time-scale since the interaction } of \textcolor{black}{$\sim$ 1.2 $\times$ 10$^{8}$} yr \footnote{60\degree/360\degree\ $\times$ 0.7 $\times$ 10$^{9}$ yr = \textcolor{black}{1.2 $\times$  10$^{8}$ yr} }\textcolor{black}{; and} } (iv) \textcolor{black}{the NGC\,90 optical}  morphology agrees well with that of the  \cite{oh08} model, their figure 1,  at  \textcolor{black}{t = 0.1 and 0.2} Gyr. \textcolor{black}{While the uncertainties for each of these time-scales are large and difficult to quantify, taken together they indicate} that the interaction probably occurred within the last \textcolor{black}{$\sim$ 1.0 -- 2.5} $\times$ 10$^8$ yr. The presence of a prominent tidal tail in this time frame indicates that the  \cite{oh08} models apply to the interaction and implies an interaction with   $S <$ 0.3.\\
\textcolor{black}{The $\sim$ 1.0 -- 2.5}  $\times$ 10$^8$ yr interaction time-scale indicates  NGC\,93 is the most probable perturber of NGC\,90 which would rule out interactions with other similar sized SRGb063 galaxy group members (see Fig. 1, right panel) projected at greater distances. \\
Tidal interactions between pairs of galaxies are known to be capable of displacing substantial fractions of their \hi, but it is quite unusual for the  the highest density \hi\  \textcolor{black}{to be observed} beyond the optical disks. \textcolor{black}{Some other} examples are NGC\,3921 \citep{hibbard96},  Arp\,105 \citep{duc97} and Arp\,181 \citep{seng13}.


\hi\ masses for  galaxies of  a size and morphological type similar to NGC\,90 and NGC\,93  are expected to be  $\sim$ 1.0 $\times$   10$^{10}$ M$_\odot$ and  $\sim$ 4.3 $\times$   10$^{9}$ M$_\odot$ respectively, \textcolor{black} {with an error of $\sim$50\% }\citep{hayn84}.  If we assume that it is only the low velocity \hi\ component (see section \ref{results_content}) that  remains \textcolor{black}{viralised within the  NGG\,90 potential}, then its \hi\ mass is $\sim$ 3.4 $\times$ 10$^{9}$ M$_\odot$,  with a higher than usual uncertainty ($\ge$ 10\%) because this mass is derived from an approximate separation of the NGC\,90 spectrum into high- and low-velocity components. These complications together with the uncertainty  in  the expected \hi\ mass values \citep{hayn84} prevents us from making a conclusive statement on the H{\sc i} deficiency of NGC\,90, although it is possible that its apparent \hi\ deficiency is real.  \textcolor{black}{Based on  the marginal detection of \hi\ \textcolor{black}{ in NGC\,93 from our} GMRT map we conclude that NGC\,93 has no significant \hi\ content. } Again the expected H{\sc i} mass estimate of NGC\,93 is also highly uncertain as its morphological type is not clearly known (Table \ref{table_1}). But  marginal  GMRT detection of H{\sc i} in NGC\,93 strengthens the  case for it being  H{\sc i} deficient.

\subsection{ Possible explanations}
While it remains unclear whether the Arp\,65 pair, individually or together, is  H{\sc i} deficient, we observed a massive H{\sc i} cloud without detected  optical, \textcolor{black}{UV or  \textit{Spitzer} 24 $\mu$m} counterparts, projected beyond the pair's optical disks.   Additionally, \textcolor{black}{  the  prominent tidal arms or tails at optical wavelengths are a clear signature  of a recent tidal interaction \textcolor{black}{($\sim$ 1.0 -- 2.5}  $\times$ 10$^8$ yr).}  \textcolor{black}{We discuss three possible scenarios  below which could explain these features and reach a conclusion about } which is the most probable scenario.


\subsubsection{\textcolor{black}{Intra-group medium } and ram pressure}
 NGC\,90 is projected $\sim$  3.3 arcmin (70 kpc) east of the  intra-group medium (IGM) X--ray emission centroid of the SRGb0063 group mapped by \cite{mahdavi00}. \textcolor{black}{In \textcolor{black}{their} ROSAT map}, NGC\,90 is projected at the easternmost edge of the 3 $\sigma$  contour. The SRGb0063 group has an L$_X$ (0.5)  = 3.63 x 10$^{42}$ ergs s$^{-1}$  \citep{mahdavi04} and \textcolor{black}{a} velocity dispersion of 336 $\pm$36 \km. NGC\,90 has a V$_{opt}$ of 5197 \km, \textcolor{black}{which is}  574 \km\ lower than the group V$_{opt}$ = 5771 $\pm$48 \km\ \citep{mahdavi04}.

\textcolor{black}{ The strong ram pressure that can arise in
galaxy clusters because of  their dense ICM  and high velocity dispersions \textcolor{black}{is capable of stripping and displacing} large quantities of H{\sc i}\textcolor{black}{.  But ram pressure is predicted to be}  much weaker in galaxy groups. Comparing the  L$_X$ and velocity dispersion of the SRGb063 group with that of galaxy clusters, we see the  SRGb063 L$_X$  is about two orders of magnitude lower  and the velocity dispersion is a factor of 2 to 3 times lower than for galaxy clusters. We would therefore expect ram pressure in the SRGb063 group to be at least an order of magnitude or two lower  than in clusters. While there is evidence  that ram-pressure stripping has produced \hi\ deficiencies in galaxy groups with X-ray emitting IGM
\citep[e.g.][]{davis97, seng06}, \textcolor{black}{several studies concur that for galaxies  with M* similar to NGC\,90 \textcolor{black}{(whether in galaxy  groups or galaxy clusters)},} large offsets of their highest density \hi\ beyond optical disk  cannot  be attributed to ram pressure alone \citep[][]{bravo00,chung09,scott10}. } Still, we also note that the extent of the \textcolor{black}{NGC\,90 \hi\ disk (assuming it is traced by the low-velocity \hi\ component) is similar to that of its  optical disk}. Taken in isolation this could suggest ram-pressure stripping, but in this case the recent tidal interaction seems a more likely cause for \textcolor{black}{the truncation of low-velocity \hi\ disk}.  We conclude that while there may be low-level ram-pressure stripping, it is highly unlikely to be cause of the \textcolor{black}{displacement of} the high-density \hi\  gas to the SE of NGC\,90.

\subsubsection{ SE high-velocity \hi\ mass belongs to a dwarf galaxy}
An alternative scenario  \textcolor{black}{is} that the entire SE high-velocity \hi\ mass belongs to a small dwarf galaxy. { However a dwarf galaxy with \textcolor{black}{ an \hi\ mass of } $\sim$ 3.4 $\times$ 10$^{9}$ M$_\odot$ is  expected to have a stellar mass of least a few times 10$^{8}$ M$_\odot$ and a surface brightness of $\sim$ 24--25 magnitude arcsec$^{-2}$, which would  be visible in the SDSS and UV images. But in this case there is no visible optical or UV counterpart  at these wavelengths. Moreover the velocity field of the  SE high-velocity \hi\ mass is irregular, more resembling  debris than a galaxy.}

\subsubsection{Gas-rich NGC\,90 interacts with gas-poor NGC\,93 }
In this scenario the  SE high-velocity \hi\ mass \textcolor{black}{ that} lies between NGC\,90 and NGC\,93 in velocity space \textcolor{black}{ and  offset $\sim$ 25 kpc in projection SE} of the NGC\,90 disk (see Fig.2, bottom left panel) \textcolor{black}{ is a result of the tidal interaction between NGC\,90 and NGC\,93 ~\textcolor{black}{which also } produced the optical interaction signatures in NGC 90.} The reason for the marginal GMRT \hi\ detection in NGC\,93  is that it was already essentially devoid of \hi\ prior to the interaction, which means \textcolor{black}{ almost all of the \hi\ detected with the GMRT} belonged to NGC\,90 before the interaction.  \textcolor{black}{ Had NGC\,93 contained a significant mass of \hi\ before  the interaction, we would expect \textcolor{black}{the bulk of} it to have remained bound to NGC\,93 because of its deeper potential well compared to NGC\,90.  } In this  scenario the pair interaction caused  a massive  displacement of about half of the \hi\ in NGC\,90 (the SE high-velocity \hi\ mass) \textcolor{black}{ $\sim$ 1.0 -- 2.5  $\times$ 10$^8$ yr ago}.  \textcolor{black}{This  an  effect on \hi\ similar to the one} that caused the massive \hi\ displacement beyond the optical disks observed  in another interacting pair, Arp\,181 \citep{seng13}. But, unlike Arp\,181 (M* ratio 1:1.6), in Arp\,65 the displacement is primarily  in the velocity plane, although with some offset in the plane of the sky. The displaced \hi\ was tidally stripped from NGC\,90 during the interaction and accelerated toward the NGC\,93 velocity during the closest approach, attracted by the deeper gravitational potential of NGC\,93  (see Fig. 2, middle right panel). The $\sim$ 3.4 $\times$ 10$^{9}$ M$_\odot$ \hi\ that \textcolor{black}{is now observed within the NGG\,90 optical disk } (the low-velocity NGC\,90 \hi\ component) rapidly re-virialised in the gravitational potential of the galaxy and now presents as a rotating \hi\ disk. In contrast, despite having a similar  \hi\ mass and greater maximum \hi\ column density than the low-velocity NGC\,90 component, the SE \hi\ mass \textcolor{black}{ remains as evolving \hi\ debris that currently lacks \hi\ rotation } and SF signatures. \textcolor{black}{Assuming NGC\,93  orbit takes it farther from NGC\,90 before their next encounter it seems likely that the near-term fate of the  SE \hi\ mass will be to fall back into the NGC\,90 potential}.  While it would take tailored modelling to prove  this scenario, it appears to provide the most reasonable explanation amongst the alternatives considered  for observations.

\subsection{Star formation in the tidal debris }

The  FUV (\textit{GALEX}) image of the Arp\,65 pair reveals several interaction-induced star-forming zones in the tidal arms and outer disk of NGC\,90. At the base of the tidal tail projected \textcolor{black}{north-west} of the NGC\,90 optical disk, a particularly bright hinge clump was reported by \cite{hancock09} (labelled as SF1 in Fig. 2). This clump shows strong emission in \textit{GALEX} FUV and NUV images and is very bright in the  \textit{Spitzer} 8$\mu$\textcolor{black}{m} image (Fig. 1 in  \cite{hancock09}). In addition, beads of star formation are also visible all along the NGC\,90 tail north-west of the hinge clump. There are several star-forming zones to the east and south-east of  NGC\,90,  and several of them are beyond the  optical disk (the prominent ones  are labelled as SF2, SF3  in Fig. 2, top panel). We derived the stellar masses of these clumps using   online \textit{Spitzer } IRAC 3.6$\mu$\textcolor{black}{m} and 4.5$\mu$\textcolor{black}{m} photometry \textcolor{black}{data} \citep{smith07} and the recipe from \cite{eskew12}.  The hinge clump (SF1) is the most massive star-forming zone with an estimated stellar mass $\sim$ 4.5 $\times$ 10$^{8}$ M$_{\sun}$. \textcolor{black}{The stellar mass estimates for SF2 and SF3 } are   1.5 $\times$ 10$^{8}$ M$_{\sun}$ and 7.2 $\times$ 10$^{7}$ M$_{\odot}$ respectively. Uncertainties \textcolor{black}{for} these masses are $\sim$ 10\%. \textcolor{black}{In the \textit{Spitzer}  images SF3 consists of at least two clumps while SF2 } has a bright main clump and a diffuse area of star formation around it.  \textcolor{black}{ The  star formation rates (SFR) of each clump were estimated using} \textit{Spitzer} MIPS 24 $\mu$\textcolor{black}{m} data using the  calibration reported by \cite{rieke09}.  As expected, the stellar clump with the highest mass, SF1, shows the highest SFR amongst the three main star-forming zones. The SFRs of SF1, SF2 and SF3 were estimated to be 0.025, 0.003 and 0.007 M$_{\odot}$ yr$^{-1}$, respectively.


The H{\sc i} image of NGC\,90 (Fig. 2, top panel) reveals that all three star formation zones are embedded in  H{\sc i} tails and debris. However, the  \hi\ column densities at the positions of the SF zones vary between 1.2 -- 4.7 $\times$ 10$^{20}$ cm$^{-2}$. None of the three SF zones coincide with the local maxima of the  H{\sc i} column density.  The highest H{\sc i} column density in these regions  is $\sim$  4.7 $\times$ 10$^{20}$ cm$^{-2}$, where the synthesised beam samples \textcolor{black}{a} 16.6 kpc $\times$ 13.2 kpc area.  The high-resolution \textcolor{black}{\hi\  maps were} made using \textcolor{black}{a taper on the} \textit{u,v} data and also yielded similar results. The bottom right panel of Fig. 2  shows the last three highest column density contours (in black), between 1.5 -- 8.3 $\times$ 10$^{20}$ cm$^{-2}$, \textcolor{black}{ overlaid on the low-resolution H{\sc i} contours (in white)} and the SDSS optical image of NGC\,90. Only the high column density contours are plotted to \textcolor{black}{highlight} the H{\sc i} local maxima.  Here the synthesised beam samples an area of 8.0 kpc $\times$ 5.2 kpc. The distribution appears to be more clumpy and fragmented than its low-resolution counterpart, however none of the H{\sc i} clumps coincide with the known star-forming zones. While there is no  consensus in the literature about the threshold H{\sc i} column density at which  star clusters form in tidal  debris and outer disk regions, \cite{maybhate07} found an  increase in the number of star clusters above N(H{\sc i}) $\sim$ 4 $\times$ 10$^{20}$ cm$^{-2}$. They suggested that a  H{\sc i} column density  threshold may be a necessary but not a sufficient condition for star cluster formation. An important point to be noted here is that the synthesised beam size plays a crucial role and the largest spatial scale sampled in \cite{maybhate07} was 7.5 kpc.  \textcolor{black}{Our  high- and low-resolution GMRT maps sample between 7 kpc to 16 kpc respectively}.  Considering this, we find that in NGC\,90, most of the star-forming zones more or less  satisfy the threshold of \cite{maybhate07}.  Moreover, as suggested in \cite{maybhate07}, in Arp\,65  we also find that this threshold  is indeed not a sufficient criterion for star formation.   The highest column density  \hi\ in Arp\,65  ($\sim$ 8.0 $\times$ 10$^{20}$ cm$^{-2}$, in our high-resolution maps) was detected  within SE high-velocity \hi\ mass.  

\textcolor{black}{ A star formation map of NGC\,90 (Fig. 4) was created by co-adding \textit{Spitzer} 24 $\mu$m  and \textit{GALEX}  images in units of M$_{\sun}$ yr$^{-1}$ pix$^{-1}$ where the size of each pixel is 2.4". We used sky-background-subtracted archive image and converted the observed flux to
  star formation rates using the calibrations provided by \cite{rieke09} and \cite{Kennicutt98} for  24 $\mu$m and FUV emission, respectively. The figure  shows no sign of any star-forming activity above  the sky-noise limit  within a radius of 20" ($\sim$ 7 kpc) of  the highest \hi\ column density  of the SE high-velocity \hi\ mass. Sky background subtraction for the \textit{Spitzer} 24 $\mu$m image  was difficult.  The archival \textit{Spitzer} 24 $\mu$m images are not cleaned properly and the background due to CCD reflection is severe. The observed typical rms across the map is $\sim$2 $\times$ 10$^{-5}$ M$_{\odot}$yr$^{-1}$ per pixel and can be considered as a reasonable upper limit. Using this,  the integrated upper limit SFR within a 20$^{\prime\prime}$ radius of the H{\sc i} column density maximum in the SE high-velocity \hi\ mass is 2.2  M$_{\odot}$yr$^{-1}$. No evidence of star formation associated with this high \hi\ column density was found in  the SDSS, \textit{Spitzer} or \textit{GALEX} data.}

\section{Concluding remarks}
\label{concl}
We have mapped the \hi\ in Arp\,65 (NGC\,90 and NGC\,93) with the GMRT and found that approximately half  of the detected \hi\ is in a high-velocity component projected SE of the NGC\,90 optical disk, which was probably removed  from NGC\,90 during the interaction  \textcolor{black}{ $\sim$ 1.0 -- 2.5  $\times$ 10$^8$ yr ago} with its more massive gas-poor partner NGC\,93. The balance of the  \hi\ is detected in a disk rotating at a lower velocity that is associated with NGC\,90. The  high-velocity \hi\ component contains the H{\sc i} column density maxima  $\sim$  1.2 arcmin (25 kpc) E  of  the NGC\,90 optical centre. The  tidal interaction that produced the impressive optical tidal tails in NGC\,90 was probably also responsible for removal of \hi\ from the NGC\,90 disk.  This means that  the asymmetric \hi\  morphology of NGC\,90 disk is the result of the superposition of gas removed from NGC\,90 during its tidal interaction with NGC\,93 on the remaining NGC\,90 \hi\ disk.  Our observations of Arp\,65 and Arp\,181 indicate  that  high-impact  pre-merger tidal interactions  involving large spirals, with M* ratios in the range $\sim$ 1:1.6 to 1:3,  have the ability to displace large factions of the \hi\ beyond the optical disks of the participants. Arp\,65 belongs to \textcolor{black}{ a small number of galaxy pairs where the highest density \hi\ is observed beyond the optical disks  of the pair following an  interaction}  \citep[e.g.][]{hibbard96, seng13}. We found that NGC\,90 has extended \textcolor{black}{on-going star formation   in the outer optical disk,} most likely \textcolor{black}{ triggered} by interaction. In the three main star-forming clumps in the tidal debris and arm, we found that the star formation is occurring in regions with \hi\ column densities  $\sim$  4.7 $\times$ 10$^{20}$ cm$^{-2}$ or lower.  But in contrast to Arp\,181, which contains a star-forming  tidal dwarf candidate,  no signature of star formation was found near the highest  column density \hi\ located in the high-velocity \hi\ debris south-east of the NGC\,90 optical disk. This debris hosts no star formation \textcolor{black}{ signatures } and reaffirms that high H{\sc i} column densities \textcolor{black}{are a} necessary, but not  sufficient criterion for star formation.



\section {Acknowledgements}
We are grateful to the anonymous referee for the helpful
remarks that have significantly improved the paper. We thank the staff of the GMRT who have made these observations
  possible. GMRT is run by the National Centre for Radio Astrophysics
  of the Tata Institute of Fundamental Research.
The Nasa Extragalactic Database, NED, is operated by the Jet Propulsion Laboratory, California Institute of Technology, under contract with the National Aeronautics and Space Administration.
We acknowledge the usage of the HyperLeda database (http://leda.univ-lyon1.fr).
Funding for the SDSS and SDSS-II has been provided by the Alfred P. Sloan Foundation, the Participating Institutions, the National Science Foundation, the U.S. Department of Energy, the National Aeronautics and Space Administration, the Japanese Monbukagakusho, the Max Planck Society, and the Higher Education Funding Council for England. The SDSS Web Site is http://www.sdss.org/. The SDSS is managed by the Astrophysical Research Consortium for the Participating Institutions. The Participating
– 13 –
Institutions are the American Museum of Natural History, Astrophysical Institute Potsdam, University of Basel, University of Cambridge, Case Western Reserve University, University of Chicago, Drexel University, Fermilab, the Institute for Advanced Study, the Japan Participation Group, Johns Hopkins University, the Joint Institute for Nuclear Astrophysics, the Kavli Institute for Particle Astrophysics and Cosmology, the Korean Scientist Group, the Chinese Academy of Sciences (LAMOST), Los Alamos National Laboratory, the Max-Planck-Institute for Astronomy (MPIA), the Max-Planck-Institute for Astrophysics (MPA), New Mexico State University, Ohio State University, University of Pittsburgh, University of Portsmouth, Princeton University, the United States Naval Observatory, and the University of Washington.

\bibliographystyle{aa} 
\bibliography{cig}

\newpage
\onecolumn

\begin{figure*}
\centering
\includegraphics[scale=0.5]{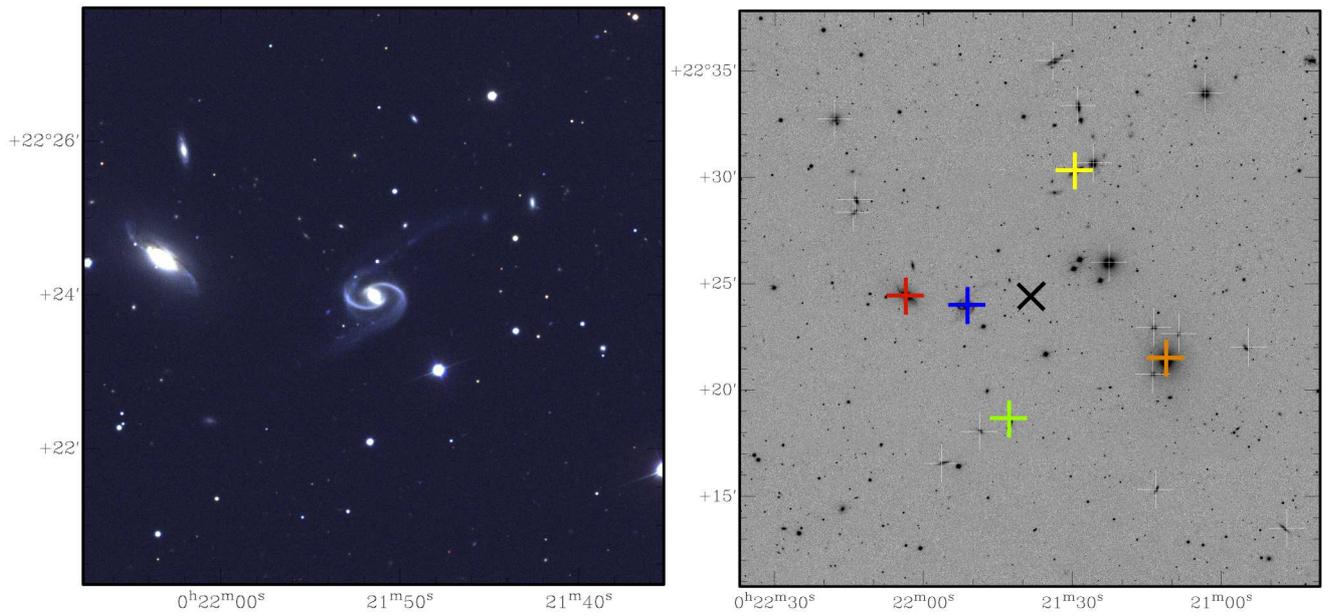}  
\caption{ \textbf{Left: Arp\,65 pair:} SDSS \textit{g,r,i}-band false-colour image. The  galaxy at the centre of the image is NGC\,90 and the galaxy of   similar size at the eastern edge of the image is NGC\,93. \textbf{Right: SRGb063 group}: \textcolor{black}{SDSS \textit{r} -- band image of the Arp\,65 pair in the  SRGb063 group. The blue and the red crosses are NGC\,90 and NGC\,93 respectively. Crosses with other colours represent nearby massive group members: CGCG 479-012 (green), NGC\,85B (yellow), NGC\,80 (orange). The black X marks the X--ray centroid of SRGb063 group from \cite{mahdavi00}.    }}
\label{fig1}
\end{figure*}

\begin{figure*}
\centering
\includegraphics[scale=0.85]{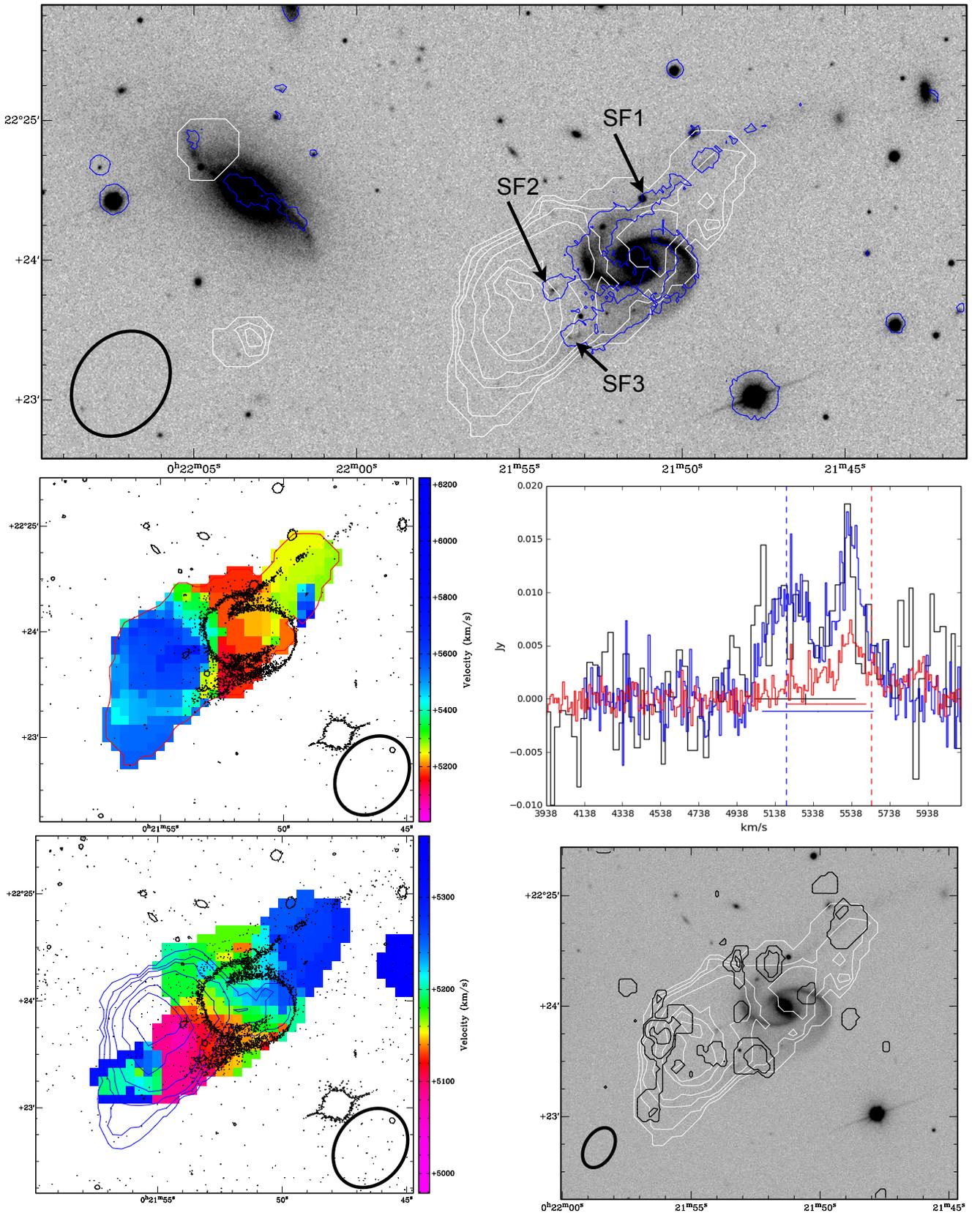}  
\caption{\textcolor{black}{\textbf{Arp\,65 Top:}Integrated \hi\  intensity contours (white) on SDSS r-band image. The blue contours are FUV from \textit{GALEX}.  The H{\sc i} column density levels are 10$^{20}$ cm$^{-2}$ $\times$  (0.6, 1.2, 1.8, 2.9, 4.7, 5.9, 7.2)  \textbf{Middle left:} The \hi\ velocity field for NGC\,90, in colour. The black contour shows the  \textit{GALEX} FUV extent NGC\,90. The red contour is the H{\sc i}  0.6 $\times$ 10$^{20}$ cm$^{-2}$ contour from the top panel. \textbf{Middle right:} H{\sc i} spectra for NGC\,90: GMRT (black) and  single dish (blue) from \cite{springob05}. NGC\,93: single dish  spectrum (red) \textcolor{black}{ \cite{springob05} spectrum for} NGC\,93. The optical velocities of NGC\,90 and NGC\,93 are marked with dashed lines.} \textbf{Bottom left: } Velocity field of the low-velocity component, overlaid with  of the total intensity contours of the high-velocity component (blue) and FUV  contours (black) define the edge of the galaxy. \textbf{Bottom right:} High-\textcolor{black}{resolution} \hi\ intensity contours (black) overlaid on low-resolution \hi\ intensity contours (white) and SDSS r -- band image. The low-resolution contours are same as the top panel. The high-resolution contours are 10$^{20}$ cm$^{-2}$ $\times$  (1.5, 4.6, 8.3). \textcolor{black}{For each of the maps the beam  size  is shown by a black ellipse at the bottom of each map.}}  
\label{fig2}
\end{figure*}


\begin{figure*}
\centering
\includegraphics[scale=0.45]{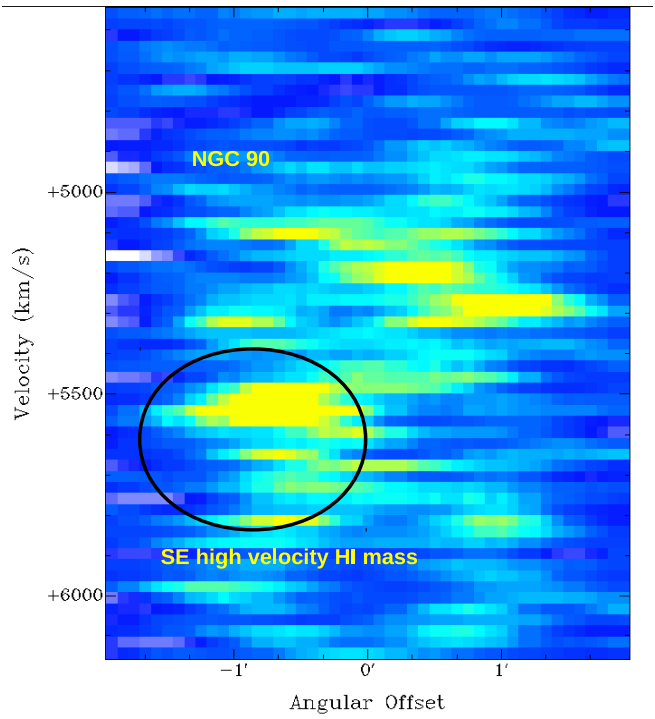}  
\caption{Position-velocity (PV)  diagram of the \hi\ mass detected with NGC 90. Both the regular disk of NGC 90 and the SE high-velocity \hi\ mass (within the black ellipse) are visible in the PV diagram.   }  
\label{fig6}
\end{figure*}

\begin{figure*}
\centering
\includegraphics[scale=0.45]{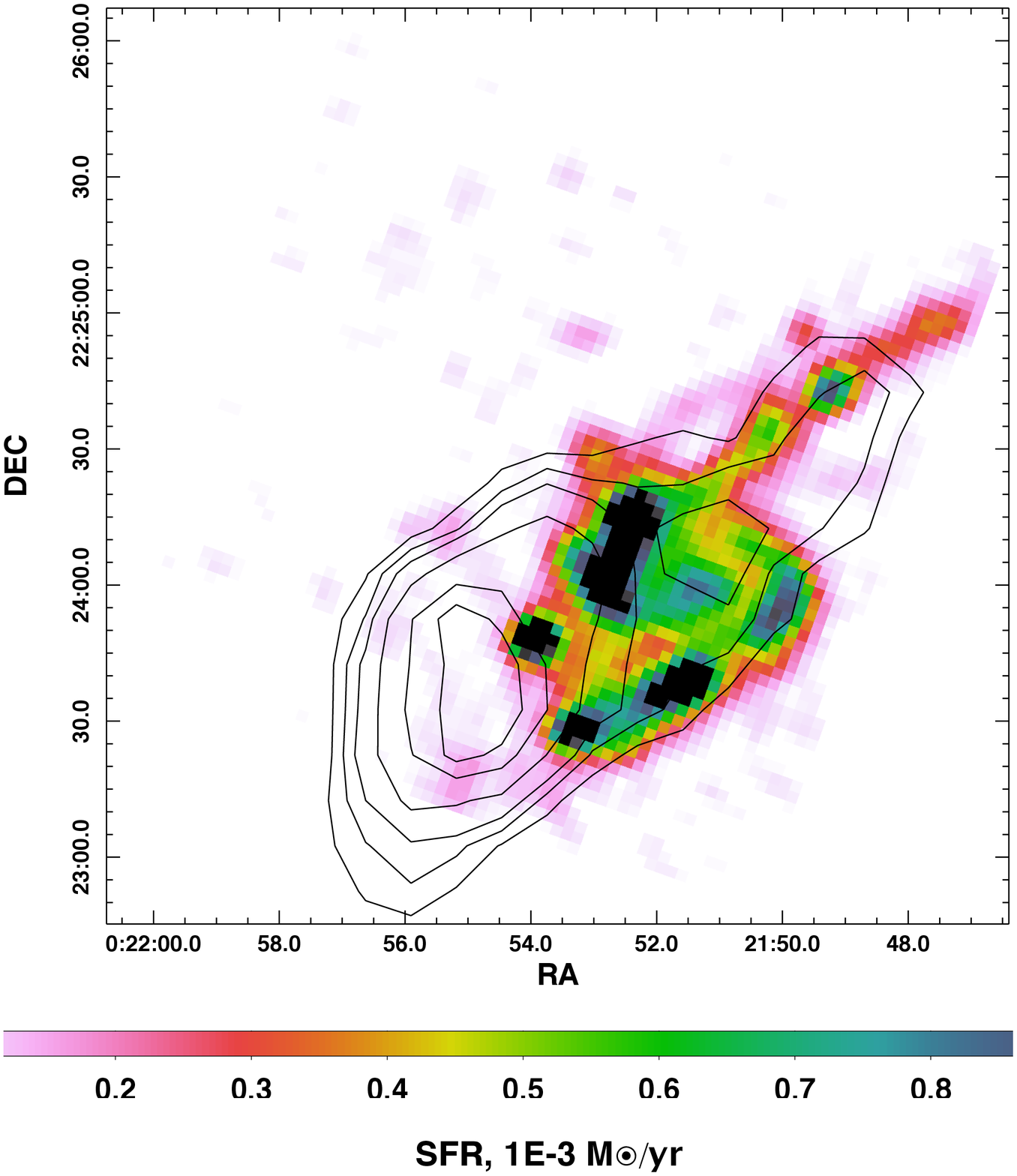}  
\caption{\textcolor{black}{ \textit{Spitzer} 24 $\mu$\textcolor{black}{m} + \textit{GALEX} FUV star formation map of \textcolor{black}{NGC\,90} . Low-resolution integrated \hi\ intensity contours (same as Figure 2) are also plotted on the star formation map.}   }  
\label{fig5}
\end{figure*}


\newpage

\begin{sidewaysfigure}
\centering
\includegraphics[scale=1.15, angle=-90]{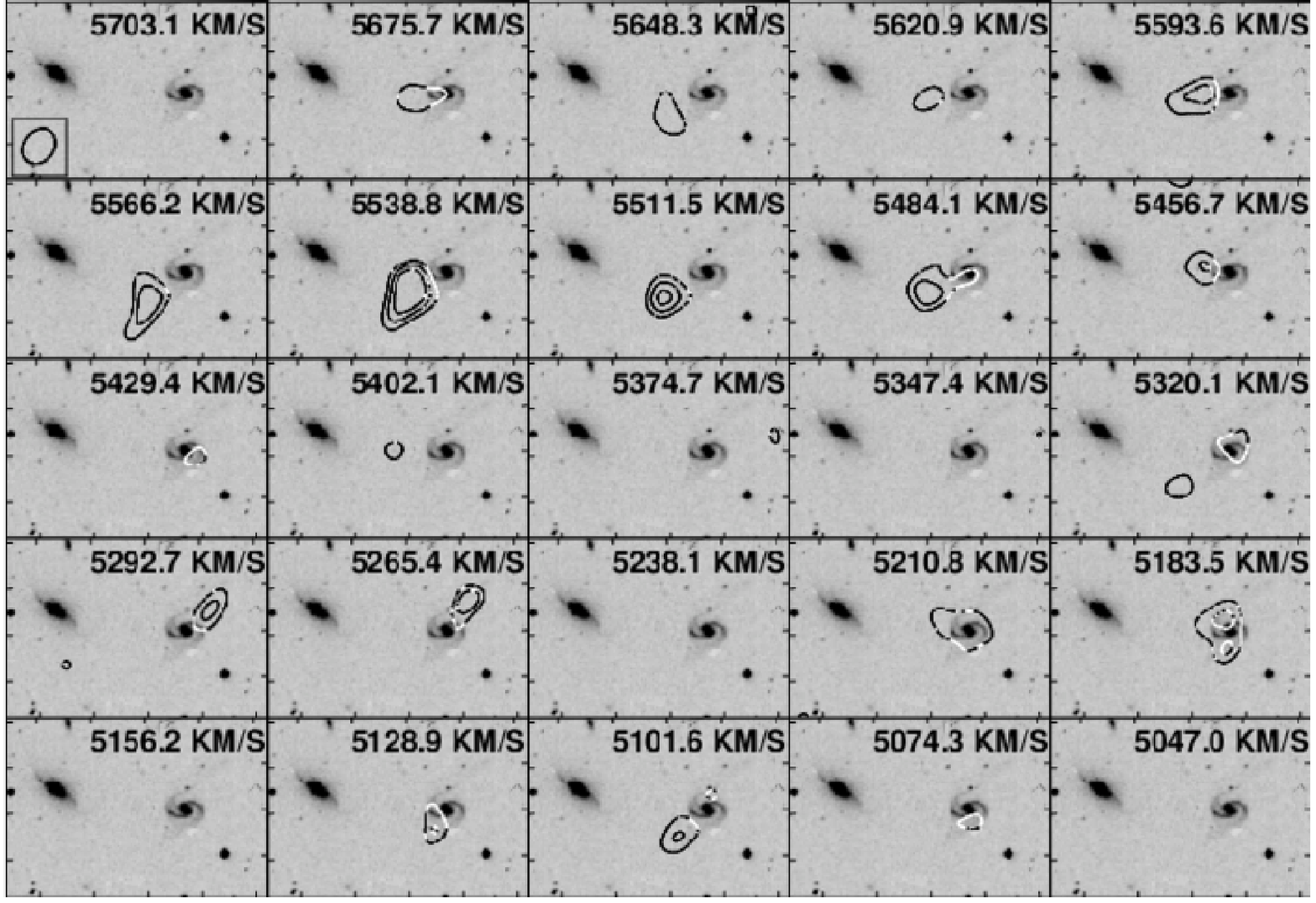}  
\caption{H{\sc i} channel images for the Arp\,65 pair. The contour levels plotted are 1.2 mJy (3,4,5). The beam (48.32 $^{\prime\prime}$ $\times$ 38.29 $^{\prime\prime}$) is shown in the bottom left corner of the topmost channel image. } 
\label{fig3}
\end{sidewaysfigure}


\end{document}